\documentclass[epj]{svjour}

\usepackage{graphicx}
\usepackage{amsfonts}
\usepackage{amsmath,nccmath}
\usepackage[colorlinks, linkcolor=blue, citecolor=blue, urlcolor=blue, breaklinks=true]{hyperref}

\newcommand{\cl}[1]{\mathcal{#1}} 
\newcommand{\on}[1]{#1^\dag \, #1} 
\newcommand{\comm}[2]{[#1 \, , \, #2 ]} 
\newcommand{\ev}[1]{ \langle #1 \rangle } 
\newcommand{\evN}[1]{ \left \langle #1 \right \rangle } 
\newcommand{\szim}[1] { \{ #1 \}_{s} } 
\newcommand{\opvec}[1]{\hat{\textbf{#1}}}
\newcommand{\gammavec}{\boldsymbol{\gamma}}
\newcommand{\ket}[1]{ | #1 \rangle }
\newcommand{\bra}[1]{ \langle #1 | }

\numberwithin{equation}{section}

\def \titleofpaper {Multimode mean-field model for the quantum phase transition of a Bose-Einstein condensate in an optical resonator}
\def \firstauthor {G. K\'onya}
\def \eaddress {domokos@szfki.hu}

\begin{document}

\title{\titleofpaper}

\author{G. K\'onya\inst{1}, G. Szirmai\inst{1,2}, P. Domokos\inst{1}\thanks{Corresponding author: \eaddress}}
\institute{Research Institute for Solid State Physics and Optics, Budapest, P.O Box 49, H-1525, Hungary\and ICFO-Institut de Ci\`encies Fot\`oniques, Mediterranean Techonoly Park, 08860 Castelldefels (Barcelona), Spain}

\titlerunning{Multimode mean-field model for a BEC in a cavity}

\hypersetup{pdftitle={\titleofpaper}}
\hypersetup{pdfauthor={\firstauthor\ (\eaddress)}}

\abstract{We develop a mean-field model describing the Hamiltonian interaction of ultracold atoms and the optical field in a cavity. The Bose-Einstein condensate is properly defined  by means of a grand-canonical approach. The model is efficient because only the relevant excitation modes are taken into account. However, the model goes beyond the two-mode subspace necessary to describe the self-organization quantum phase transition observed recently. We calculate all the second-order correlations of the coupled atom field and radiation field hybrid bosonic system, including the entanglement between the two types of fields.}

\maketitle

\section{Introduction}
\label{sec:intro}

\begin{figure}[htbp]
\begin{center}
\includegraphics[width=0.2\textwidth]{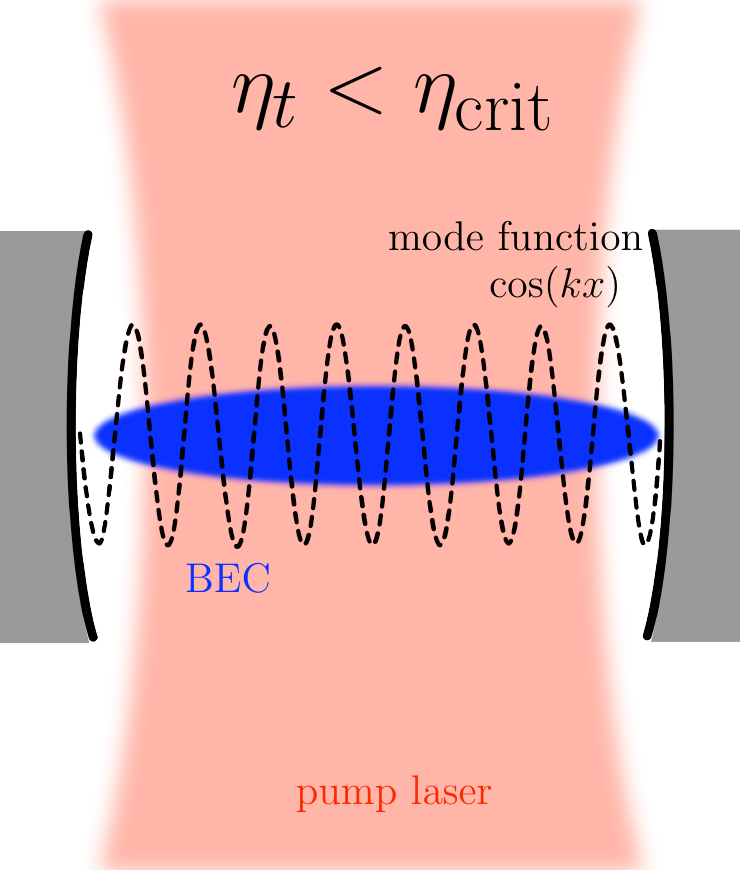}
\includegraphics[width=0.2\textwidth]{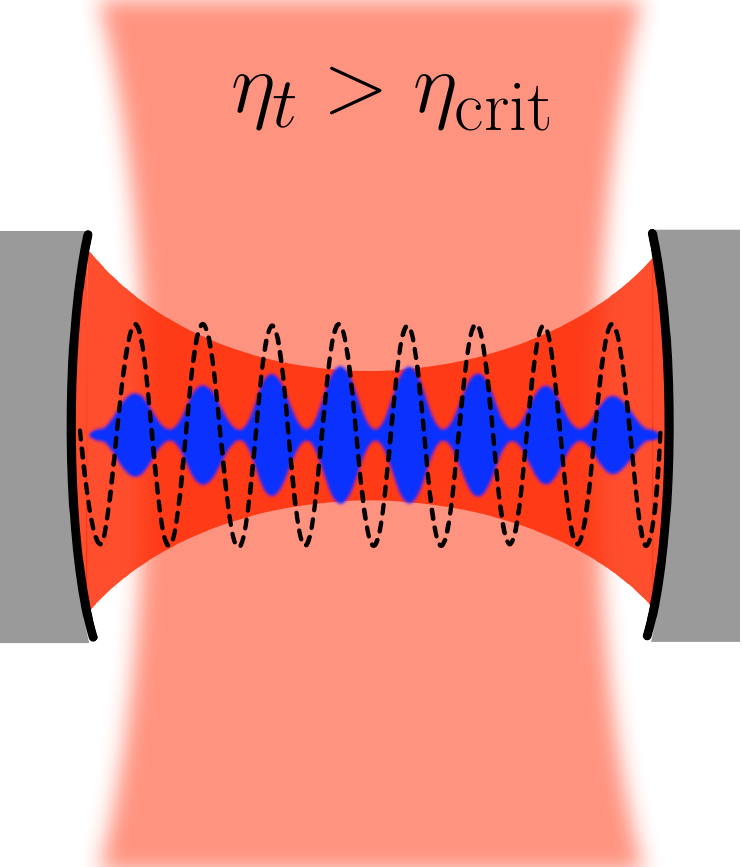}
\end{center}
\caption{Self-organization of a Bose-Einstein condensate in a cavity. Below a certain threshold pump power (left), the ultracold atoms have a quasi-homogeneous distribution, and the cavity field with mode function $\cos(kx)$ is empty. Above threshold (right), the atoms self-organize into a $\lambda$-periodic ordered lattice in which they scatter the pump light constructively into the resonator mode. There is another,  $\frac{\lambda}{2}$-shifted lattice possible to be formed.}
\label{fig:SelforgScheme}
\end{figure}

A thermal cloud of cold atoms interacting with a single mode of a high-finesse optical cavity undergoes a phase transition when tuning the power of a laser field which illuminates the atoms from a direction perpendicular to the cavity axis. Below a threshold pump power, the cloud is homogeneous which is stabilized by thermal fluctuations. In this phase, the optical mean field in the cavity is zero, because the laser pump field  is not scattered into the cavity mode from the homogeneous distribution of atoms.  Above a threshold pump power, however, this solution becomes unstable. Then the atoms self-organize into a wavelength-periodic crystalline order which gives rise to Bragg-scattering from the transverse pump laser into the cavity. The resulting cavity field traps the atoms in the optical lattice distribution (see Figure \ref{fig:SelforgScheme}). The self-organization effect has been first predicted in Ref.~\cite{Domokos2002Collective}, and soon experimentally observed \cite{Black2003Observation}. It is closely related to the collective atomic recoil lasing transition \cite{Javaloyes2004Selfgenerated} which is a Kuramoto-model-like synchronization phenomenon \cite{Javaloyes2008Collective}. A classical mean-field description of self-organization has been presented in \cite{Asboth2005Selforganization,Nagy2006Selforganization}, which relies on a self-consistent canonical distribution of the atomic ensemble at a finite temperature $T$. The mean-field theory in \cite{Griesser2010Vlasov} introduces a Vlasov-type equation in phase space which can account for arbitrary velocity distributions. 

The same self-organization effect can occur in the case of Bose-Einstein condensed ultra-cold atoms (BEC) at zero temperature \cite{Nagy2008Selforganization,Vukics2007Microscopic,Keeling2010Collective,Vidal2010Quantum,Larson2010SelforgDoubleWell}. For low pump power, instead of the thermal fluctuations, the homogeneous phase is stabilized by  atom-atom collisions, or, in the lack of collisions, ultimately by the zero-point kinetic energy. A sharp transition threshold is thus expected at zero temperature, too. This has been experimentally evidenced recently \cite{Baumann2010Dicke}. Moreover, in Ref.~\cite{Baumann2010Dicke} as well as in Ref.~\cite{Nagy2010DickeModel}, an analogy of the self-organization to the celebrated Dicke-model phase transition \cite{Dicke1954Coherence,Dimer2007DickeCQED,Li2010JCDicke} has been pointed out.  The dynamically coupled BEC wavefunction and single-mode cavity field realizes the Dicke-model with tunable parameters in the kHz range, and this is the first system in which the critical point can be reached and investigated.

The Dicke-type phase transition is considered usually in systems with a fixed number of atoms and where the individual atomic degrees of freedom span only a 2-mode Hilbert space. These assumptions are necessary to introduce the spin representation of the two-mode boson field. The drawback of this approach is that the role of higher excited motional modes cannot be included in the description. 

In this paper we resort to a different approach  which allows for the generalization to a multimode treatment of the matter wave field. Instead of fixing the number of atoms in the atomic modes, we invoke the grand canonical ensemble description, on assuming that the constant $N_c$ is the mean number of atoms in the condensate. In contrast to Ref.~\cite{Nagy2008Selforganization} where the condensate mean field has been written in position space, here both the condensate and the quantum fluctuations will be treated in momentum space. This is the most economic approach in terms of computational needs, since the calculation  converges very fast to the exact result as one includes higher excited kinetic energy eigenstates. We will show that the position of the critical point is not affected by the higher modes. Furthermore we will confirm that the two mode model correctly describes the system and the phase transition below, and in the vicinity of the critical point. Far above threshold, however, the effect of the higher modes will become significant and the new multimode treatment is required.  

The paper is organized as follows.  In section \ref{sec:theoretical_description}, we describe the microscopic Hamiltonian model of our system. 
In section \ref{sec:beccav_mode_expansion}, the one-particle wavefunctions and the mode expansion are introduced. The backbone of our paper is section \ref{sec:mean-field approx}, where we introduce the grand canonical Hamiltonian, which makes possible to  systematically define, in subsection \ref{subsec:mean_field}, the mean-field approximation in a multimode model.  Then, in \ref{subsec:fluctuations}, we determine the independent quasiparticles by means of a  Bogoliubov transformation.  The subsection \ref{subsec:normal_phase_fluctuations} is devoted to studying the fluctuations in the normal phase of the system (below threshold). In section \ref{sec:ground_state_analysis}, the ground state of the fluctuation Hamiltonian is analysed  and the incoherent populations in the excited modes above the condensate are calculated numerically. We show that the ground state is an entangled one of the bipartite system of  
the cavity mode and the atomic motional degrees of freedom, and the entanglement is quantified in section \ref{sec:ground_state_entanglement}.

\section{Microscopic model of the system} 
\label{sec:theoretical_description}

\begin{figure}[htbp]
\begin{center}
\includegraphics[width=0.33\textwidth]{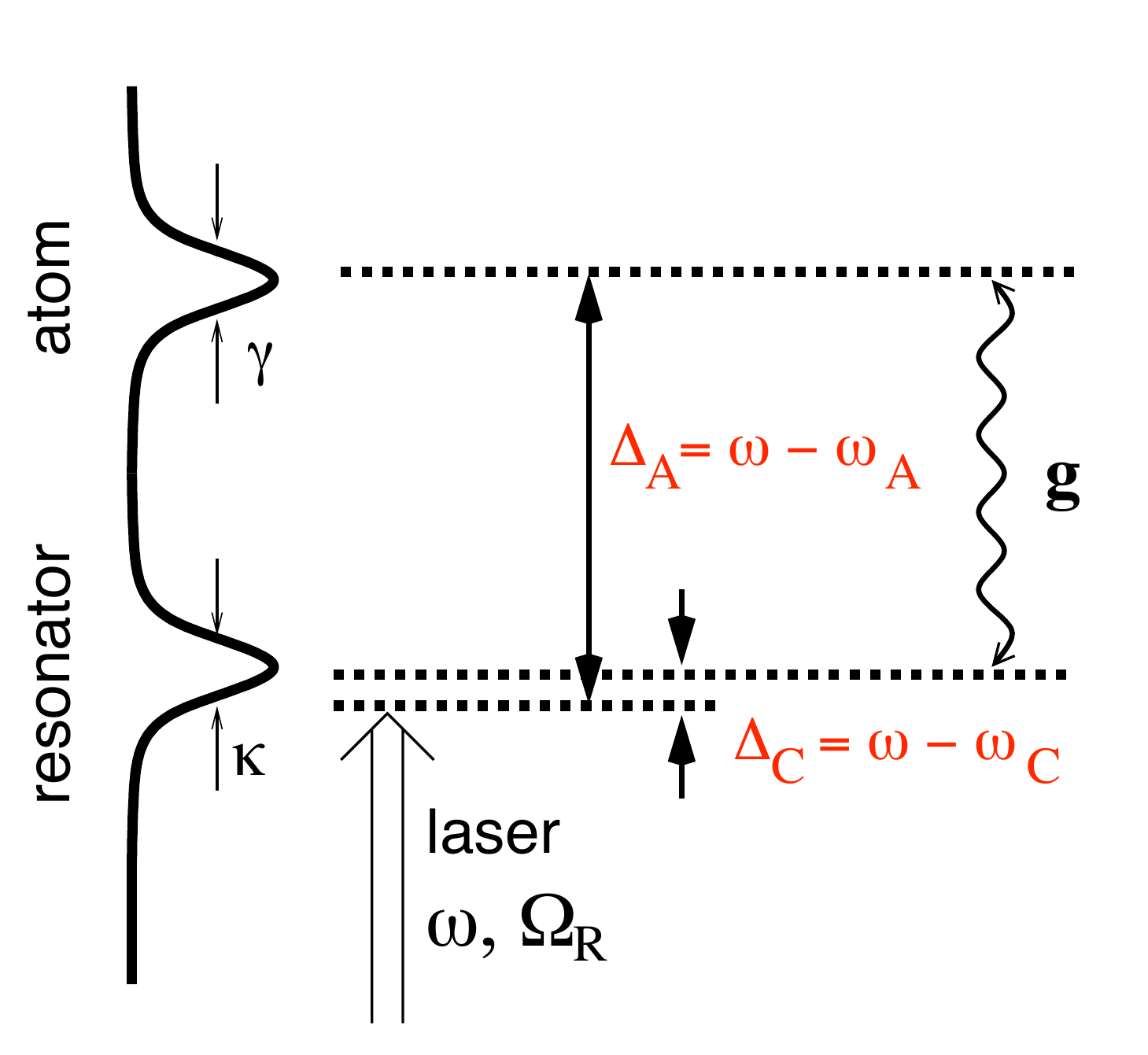}
\end{center}
\caption{Summary of the system parameters used in the microscopic model.}
\label{fig:parameters}
\end{figure}

We consider an ensemble of ultracold atoms at $T=0$ interacting with a single-mode of a high-Q optical cavity \cite{Brennecke2007Cavity,Colombe2007Strong}. The atoms are coherently driven from the side by a laser field with frequency $\omega$, directed perpendicularly to the cavity axis (see Fig.~\ref{fig:SelforgScheme}). The driving strength is described by the Rabi-frequency $\Omega_{R}$. The laser is detuned far below the atomic transition $\omega_A$, that is,  $|\Delta_A| \gg \gamma$, where $2\gamma$ is the full atomic linewidth at half maximum and the (red) atom-pump detuning is $\Delta_A=\omega - \omega_A<0$. This condition ensures that the electronic excitation is extremely low in the atoms, hence the spontaneous photon emission is suppressed.  At the same time, the laser field is nearly resonant with the cavity mode frequency $\omega_C$, i.e.\ $|\Delta_C| \sim \kappa$, where $\kappa$ is the cavity mode linewidth and the cavity-pump detuning is $\Delta_C= \omega - \omega_C$  (all these parameters are summarized in Fig.~\ref{fig:parameters}). 
The scattering of laser photons into the cavity is thus a quasi-resonant process. Moreover, it is significantly enhanced by the strong dipole coupling between the atoms and the mode due to the small  volume of the cavity. This coupling strength is characterized by the single-photon Rabi frequency $g$, which is in the range of $\kappa$. 

For the sake of simplicity, we describe the dynamics in one dimension $x$ along the cavity axis, in which direction the mode function is $\cos(kx)$, and the cavity length is $L$.  The atom field and the resonator mode are described by the pair of bosonic annihilation and creation field operators, $\hat{\Psi} (x)$, $\hat{\Psi}^\dagger (x)$ 
and $\hat{a}$, $\hat{a}^\dagger$, respectively. In the large detuning limit, i.e., $\Delta_A$ is the far largest frequency in the system, the excited state can be eliminated \cite{Moore1999Quantum}, and  the many-particle Hamiltonian in the reference frame rotating at the pump frequency $\omega$ is ($\hbar=1$) :
\begin{multline} \label{eq:beccav_H_def}
\hat{H}=-\Delta_{C} \, \on{\hat{a}} \,
+ \int_{-\frac{L}{2}}^{+\frac{L}{2}}
\hat{\Psi}^{\dagger}(x) \Bigl( -\frac{1}{2m} \frac{d^{2}}{d x^{2}} \\
+U_{0} \, \on{\hat{a}} \, \cos^{2} (kx)
+ \eta_t \, \cos(kx) \, \left( \hat{a}^\dag + \hat{a} \right) \Bigr) \hat{\Psi}(x) \, dx \,.
\end{multline}
The first term gives the optical field energy in the resonator. The second term is the kinetic energy of the atoms. The third term describes the dispersive interaction between the atoms and the cavity with a coupling strength  $U_0= g^2 /\Delta_A$. The underlying physical process is the absorption and stimulated emission of a cavity photon. This scattering process,  from the point of view of the atoms,  means an effective potential of the shape $\cos^{2} (kx)$ with depth depending on the photon number $\on{\hat{a}}$.  According to the red atomic detuning ($\Delta_A<0$) the coefficient $U_0<0$, which implies trapping positions at the $\lambda/2$ separated antinodes of the cavity mode function.  On the other hand, from the viewpoint of the cavity field, this term preserves the photon number but gives rise to a  frequency shift depending on the spatial distribution of the atom field. This term is responsible for the optomechanical-type coupling \cite{Szirmai2010Quantum,Chen2010Optomech,Bhattacherjee2010Bogoliubovnoise} investigated in the experiments \cite{Brennecke2008Cavity,Murch2008Observation}. The last term describes the effect of the pump field and results from the scattering between the pump laser and the cavity field. The back action of this scattering on the pump laser is neglected. It thus amounts effectively to a cavity field driving,  $\hat{a}^\dag + \hat{a}$, with the constant transverse pump amplitude $\eta_t=\Omega_{R} \, g / \Delta_A$,  and depending again on the local matter wave field density.

In Eq. \eqref{eq:beccav_H_def} we consider only one type of atom-atom interaction, namely the one which is mediated by the cavity field. This interaction is long-range \cite{Asboth2004correlatedmotion} and has strict periodicity due to the fixed momentum of the exchanged photons. In contrast, the effect of s-wave scattering can take place at arbitrary momentum values and causes a broadening of the atomic momentum distribution around the peaks at integer times the momentum of the cavity photon \cite{Zhang2009discretemode}. We disregard this broadening effect by assuming that the cavity mediated long range interaction dominates over s-wave scattering. Such an assumption is physically sound for a wide range of experimental paramaters \cite{Szirmai2010Quantum} and helps the distillation of the effect caused by the cavity photons.

Here we are interested in the ground state properties and the excitation spectrum of this Hamiltonian, and therefore disregard the effects of the photon leakage out of the cavity. 

\section{Mode expansion}
\label{sec:beccav_mode_expansion}

For the decomposition of $\hat{\Psi} (x)$, we can use the complete orthonormal set of mode functions, 
\begin{equation*}
\sqrt{\frac{1}{L}} \, , \, \left\{\sqrt{\frac{2}{L}} \, \cos(nkx) \right\}_{n=1}^{\infty} \, , \, \left\{\sqrt{\frac{2}{L}} \, \sin(nkx)\right\}_{n=1}^{\infty}  \,.
\end{equation*}
However, the $\sin(nkx)$ modes are not populated by the parity-conserving Hamiltonian \eqref{eq:beccav_H_def} when the system starts from a homogeneous BEC (being the ground state for $U_0=\eta_t=0$). Thus, $\hat{\Psi} (x)$ can be expanded in terms of the even modes as
\begin{equation} \label{eq:psi_Fourier_kif}
\hat{\Psi} (x)= \sqrt{\frac{1}{L}} \; \hat{c}_0 + \sum_{n=1}^{\infty} \; \sqrt{\frac{2}{L}} \; \cos(nkx) \; \hat{c}_n \; , 
\end{equation}
where the $\hat{c}_n$ operators satisfy bosonic commutation relations. We will use the compact notations $\opvec{c}= \left(\hat{c}_0 \, , \, \hat{c}_1 \, ,  \, \ldots \right)^T$, being a column vector, and $\opvec{c}^\dag= \left(\hat{c}_0^\dag \, , \, \hat{c}_1^\dag \, , \, \ldots \right)$,  being a row vector. 
Then the operator of the total atom number is:
\begin{equation} \label{eq:N_def}
\hat{N}=\int_{-\frac{L}{2}}^{+\frac{L}{2}} \hat{\Psi}^\dag (x) \, \hat{\Psi} (x) \, dx = \sum_{n=0}^{\infty} \, \hat{c}_n^\dag \, \hat{c}_n = \opvec{c}^\dag \, \opvec{c}
\end{equation}
On substituting the expansion \eqref{eq:psi_Fourier_kif} into \eqref{eq:beccav_H_def}, the Hamiltonian can be constructed in terms of \emph{quadratic forms} as
\begin{equation} \label{eq:H_modus_kif_kvadr_forma}
\begin{split}
\hat{H}= 
&-\Delta_{C} \, \on{\hat{a}} + \omega_R \, \left( \opvec{c}^\dag \, \textbf{M}^{(0)} \, \opvec{c} \right) \\
&+ \frac{\sqrt{2}}{2} \eta_t \left( \hat{a}^\dag + \hat{a} \right) \left( \opvec{c}^\dag \, \textbf{M}^{(1)} \, \opvec{c} \right) \\
&+\frac{1}{4} \, U_0 \, \on{\hat{a}} \, \left( \opvec{c}^\dag \, \left( \textbf{M}^{(2)}+2 \, \textbf{I} \right) \, \opvec{c} \right) \,,
\end{split}
\end{equation}
where $\omega_R=\frac{k^2}{2m}$ is the \emph{recoil frequency}, $\textbf{I}$ is the unit matrix, and the $\textbf{M}^{(j)}$ matrices are all real and symmetric, 
\begin{subequations}
\begin{equation}
\textbf{M}^{(0)} = 
\left( {\begin{array}{*{20}{c}}
   {{0^2}} & {} & {} & {} & {} & {}  \\
   {} & {{1^2}} & {} & {} & {} & {}  \\
   {} & {} & {{2^2}} & {} & {} & {}  \\
   {} & {} & {} & {{3^2}} & {} & {}  \\
   {} & {} & {} & {} & \cdot & {}  \\
   {} & {} & {} & {} & {} & \cdot  \\
\end{array}} \right)
\end{equation}
\begin{equation}
\textbf{M}^{(1)} = 
\left( {\begin{array}{*{20}{c}}
   0 & 1 & {} & {} & {} & {}  \\
   1 & 0 & {\frac{1}{{\sqrt 2 }}} & {} & {} & {}  \\
   {} & {\frac{1}{{\sqrt 2 }}} & 0 & {\frac{1}{{\sqrt 2 }}} & {} & {}  \\
   {} & {} & {\frac{1}{{\sqrt 2 }}} & 0 & {\frac{1}{{\sqrt 2 }}} & {}  \\
   {} & {} & {} & {\frac{1}{{\sqrt 2 }}} & 0 & \cdot  \\
   {} & {} & {} & {} & \cdot & \cdot  \\
\end{array}} \right)
\end{equation}
\begin{equation}
\textbf{M}^{(2)} = 
\left( {\begin{array}{*{20}{c}}
   0 & 0 & {\sqrt 2 } & {} & {} & {}  \\
   0 & 1 & 0 & 1 & {} & {}  \\
   {\sqrt 2 } & 0 & 0 & 0 & 1 & {}  \\
   {} & 1 & 0 & 0 & 0 & \cdot  \\
   {} & {} & 1 & 0 & \cdot & \cdot  \\
   {} & {} & {} & \cdot & \cdot & \cdot  \\
\end{array}} \right)
\end{equation}
\end{subequations}
Note that the $\textbf{M}^{(j)}$ matrix is diagonal for $j=0$, tridiagonal for $j=1$, and pentadiagonal for $j=2$. 
Since the kinetic energy difference between adjacent modes increases with the mode index, there appears a natural cutoff excluding the high energy modes from the dynamics and the system effectively has only a finite number of atom field modes. In this way, this mode decomposition will result in a significantly reduced numerical effort compared to the real-space mean-field approaches in Refs.~\cite{Nagy2008Selforganization,Horak2001Dissipative}.

\section{Mean-field approximation}
\label{sec:mean-field approx} 

We assume that the atoms form a condensate which is a macroscopic mean-field $\ev{\hat{\psi} (\textbf{r},t)}$ rotating at a unique frequency corresponding to the chemical potential $\mu$. Therefore, we transform into the picture given by the \emph{grand canonical Hamiltonian} 
\begin{equation} \label{eq:K_def}
\hat{K}=\hat{H}-\mu \, \hat{N} \; ,
\end{equation}
which defines a dynamics such that the condensate mean-field is static. The chemical potential can be determined self-consistently, by assuming a fixed density of the condensate atoms $N_c/L$. 

Let us separate the mean values of the operators  $\hat{a}$ and $\opvec{c}$ from the quantum fluctuations.  The mean-field part is  denoted by $\ev{\hat{a}} =\sqrt{N_c} \, \alpha$ and $\ev{\opvec{c}}=\sqrt{N_c} \, \gammavec$, where  the $\gammavec^T \, \gammavec=1$ normalization condition fixes the number of condensed atoms to be $N_c$. Later we will justify that $\alpha$ and the elements of $\gammavec$ can be chosen real. The $\sqrt{N_c}$ multipliers are included in order to make $\alpha$ and $\gammavec$ constant in the thermodynamic limit ($N_c \rightarrow \infty$,  $L \rightarrow \infty$).
Let us displace the operators,
\begin{subequations}
\begin{gather}
\hat{a} \rightarrow \sqrt{N_c} \, \alpha + \hat{a} \,,\\
\opvec{c} \rightarrow \sqrt{N_c} \, \gammavec  + \opvec{c} \,,
\end{gather}
\end{subequations}
which is a canonical transformation. After the displacement $\langle \hat{a} \rangle=0$ and $\langle \opvec{c} \rangle=0$, so the new operators correspond to the quantum-fluctuations.
Note that the displacement breaks the $U(1)$ symmetry of the microscopic model, which is the invariance of the Hamiltonian with respect to the transformation $\hat{\Psi}(x) \rightarrow \hat{\Psi}(x) e^{-i \varphi}$ with arbitrary phase $\varphi$. The expectation values of the total photon and atom number operators can be expressed in the displaced frame as 
\begin{subequations}
\begin{gather}
\ev{\on{\hat{a}}} \rightarrow N_c \, \alpha^2 + \ev{\on{\hat{a}}} \label{eq:<ona>_elt} \\
\ev{\on{\opvec{c}}} \rightarrow N_c + \ev{\on{\opvec{c}}} \label{eq:<onc>_elt}
\end{gather}
\end{subequations}
That is, on top of the number of $N_c$ condensate atoms, there is an incoherent population of atoms outside the condensate which appears due to the atom-photon interaction.  

The terms in $\hat{K}$ after the displacement should be grouped according to the powers of $\hat{a}$ and $\opvec{c}$,
\begin{equation} \label{eq:K_sor}
\hat{K}=\hat{K}^{(0)} + \hat{K}^{(1)} + \hat{K}^{(2)} + \hat{K}^{(3)} + \hat{K}^{(4)}
\end{equation}
Let us introduce new parameters which have constant value in the thermodynamic limit  ($N_c \rightarrow \infty$,  $L \rightarrow \infty$):
\begin{subequations} \label{dc,u,y_def}
\begin{gather}
\delta_C=\Delta_C - \frac{1}{2} \, N_c \, U_0 \\
u= \frac{1}{4} \, N_c \, U_0 \\
y= \sqrt{2N_c} \, \eta_t
\end{gather}
\end{subequations}
The zeroth order term of $\hat{K}$ is a c-number, 
\begin{multline} \label{eq:K0}
K^{(0)}=N_c \Biggl(-\delta_{C} \, \alpha^2 + \omega_R \, \left( \gammavec^T \, \textbf{M}^{(0)} \, \gammavec \right) \\
+ y \, \alpha \left( \gammavec^T \, \textbf{M}^{(1)} \, \gammavec \right)
+ u \, \alpha^2 \, \left( \gammavec^T \, \textbf{M}^{(2)} \, \gammavec \right) - \mu \Biggr) \; ,
\end{multline}
so it is irrelevant to the dynamics, but gives the mean-field energy of the system. The first and second order terms read
\begin{equation} \label{eq:K1}
\begin{split}
\hat{K}^{(1)}
&= \sqrt{N_c} \, \left( \hat{a}^\dag + \hat{a} \right) \left( \Omega(\gammavec) \, \alpha +\frac{1}{2} \, y \, \gammavec^T \,  \textbf{M}^{(1)} \gammavec \right) \\
&+ \sqrt{N_c} \,\left( \opvec{c}^\dag \, \left( \textbf{M}(\alpha) -\mu \, \textbf{I} \right) \, \gammavec + \gammavec^T \, \left( \textbf{M}(\alpha) -\mu \, \textbf{I} \right) \, \opvec{c} \right) \,,
\end{split}
\end{equation}
and
\begin{equation} \label{eq:K2}
\begin{split}
\hat{K}^{(2)}
&= \Omega(\gammavec) \, \on{\hat{a}} +  \opvec{c}^\dag \, \left( \textbf{M}(\alpha)- \mu \, \textbf{I} \right) \, \opvec{c} \\
&+ \frac{1}{2} \left( \hat{a}^\dag + \hat{a} \right) \left( \opvec{c}^\dag \, \textbf{M}'(\alpha) \, \gammavec + \gammavec^T \, \textbf{M}'(\alpha) \, \opvec{c} \right) \; .
\end{split}
\end{equation}
The effective cavity resonance frequency is
\begin{subequations}
\begin{equation}
\Omega(\gammavec)=-\delta_C + u \, \gammavec^T \, \textbf{M}^{(2)} \, \gammavec  \label{eq:Omega_def} \,,
\end{equation}
and the cross-coupling of the motional modes via the cavity mean field is expressed by the matrix
\begin{equation}
\textbf{M}(\alpha)= \omega_R \, \textbf{M}^{(0)} + y \, \alpha \, \textbf{M}^{(1)} + u \, \alpha^2 \, \left( \textbf{M}^{(2)}+2 \, \textbf{I} \right)\; , \label{eq:M(alpha)_def}
\end{equation}
\end{subequations}
which is a real, symmetric matrix valued polynomial of $\alpha$. $\textbf{M}'(\alpha)$ denotes the derivative of this polynomial with respect to $\alpha$. 

It can be seen that $\hat{K}^{(j)}$ is proportional to $N_c^{\, 1-j/2}$, which implies that the third and fourth order terms disappear in the thermodynamic limit, so we disregard them. The mean-fields  $\alpha$ and $\gammavec$ are determined by the condition that, in the grand canonical Hamiltonian, the terms linear in the fluctuations $\hat{a}$ and $\opvec{c}$ collected in $\hat{K}^{(1)}$ have to vanish. The fluctuations around the mean values are described by the bilinear Hamiltonian $\hat{K}^{(2)}$. 

\subsection{The mean field solution}
\label{subsec:mean_field}

The condition $\hat{K}^{(1)}=0$ leads to the system of equations:
\begin{subequations} \label{eq:mean_sol}
\begin{gather}
\Omega(\gammavec) \, \alpha \; + \; \frac{1}{2} \, y \; \gammavec^T \,  \textbf{M}^{(1)} \gammavec =0  \;¬†.\label{eq:alpha_sol} \\
\textbf{M}(\alpha) \; \gammavec = \mu \; \gammavec  \;¬†.\label{eq:gamma_sol}
\end{gather}
\end{subequations}
These equations define a \emph{quasi eigenvalue problem}:  $\gammavec$ is the eigenvector of the matrix $\textbf{M}(\alpha)$, and the smallest eigenvalue is the chemical potential. Because the matrix is symmetric, $\mu$ will have a real value. But the value of $\alpha$, and so the matrix itself depends on $\gammavec$ through the first equation, which renders the problem to be nonlinear. It can be solved by iteration which, as a main virtue of the present approach, is a stable and fast numerical method. 

\begin{figure}[ht]
\begin{center}
\includegraphics[width=0.45\textwidth]{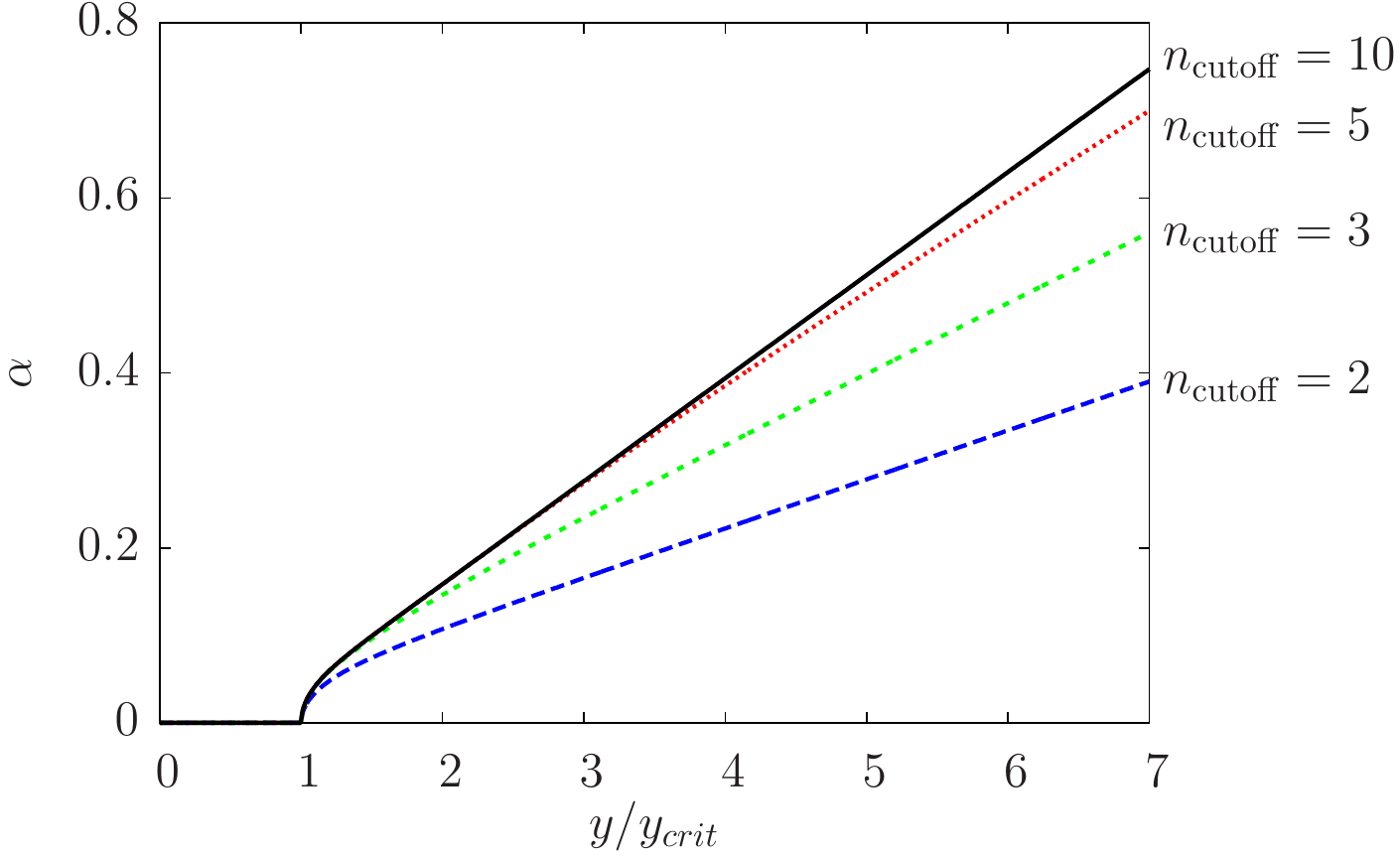}
\end{center}
\caption{The coherent field amplitude $\alpha$ as a function of the pumping strength $y$ for various cutoff mode numbers ($n_{\rm cutoff}=2 \ldots 10$). The  critical point $y_{\rm crit} = 10$ does not depend on $n_{\rm cutoff}$. The parameters: $\omega_{R}=1$, $\delta_C=-100$, $u=-20$.}
\label{fig:meanfield_alpha_y}
\end{figure}
Let us make an iteration step starting from the initial value $\alpha=0$. Then, $\textbf{M}(\alpha=0)=\omega_R \, \textbf{M}^{(0)}$, which is a diagonal matrix and its smallest eigenvalue is $\mu=0$. The corresponding normalized eigenvector is $\gammavec= \left( 1 \, , \, 0 \, , \, 0 \, , \, ... \right)^T$ for which $\Omega(\gammavec)=-\delta_C \neq 0$ and $\gammavec^T \,  \textbf{M}^{(1)} \gammavec = 0$. This yields $\alpha=0$, which is then a trivial solution describing the normal phase of our system: the resonator contains no photons and the whole condensate is in the homogeneous mode.  This solution always exists but it becomes unstable above a certain threshold $y_{\rm crit} = \sqrt{-\delta_C \, \omega_R}$. This critical point can be seen in Fig.~\ref{fig:meanfield_alpha_y} which plots $\alpha$, the mean amplitude of the cavity mode divided by $\sqrt{N_c}$, as a function of the transverse pump amplitude $y$.  This curve is calculated for various $n_{\rm cutoff}$ cutoff mode indexes, from $n_{\rm cutoff} =2$, corresponding exactly to the Dicke-model \cite{Nagy2010DickeModel}, to $n_{\rm cutoff}=10$.  The case $n_{\rm cutoff}=10$ is close to being exact, since the higher excited modes have $\gamma_n \approx 0$ with four digit precision for $n \geq 10$.
\begin{figure}[ht]
\begin{center}
\includegraphics[width=0.45\textwidth]{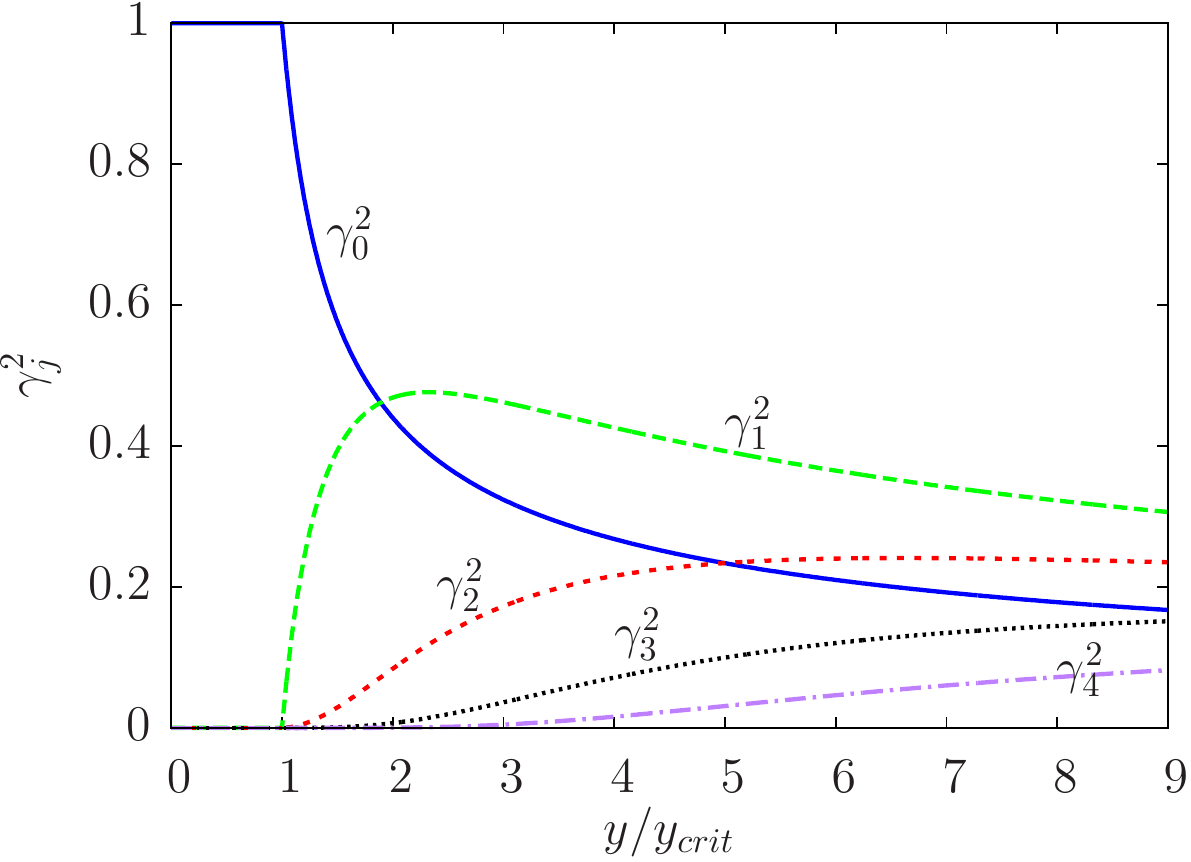}
\end{center}
\caption{The distribution $\gamma_j^2$ of condensate atoms in the modes as a function of the pump strength. The phase transition occurs in the subspace spanned by the modes $\hat{c}_0$ and $\hat{c}_1$, but well above the threshold other modes also get involved in the dynamics. The parameters: $\omega_R=1$, $\delta_C=-100$, $u=-20$, $n_{\rm cutoff}=10$.}
\label{fig:BecDistrib}
\end{figure}
The distribution of the condensate atoms in the modes is shown in Fig.~\ref{fig:BecDistrib}. Below threshold, only the homogeneous mode $n=0$ is populated. At the critical point, the population $\gamma_1^2$ in the mode $\cos{kx}$ begins to grow abruptly from zero with a finite slope, while the higher mode populations start slowly with vanishing derivative. Therefore, the two-mode approximation holds in the vicinity of the critical point. Well above threshold the other modes get populated. In Fig.~\ref{fig:meanfield_alpha_y}, above threshold, the significant dependence of the slope on the number of excited modes taken into account underlines the importance of the multimode approach as contrasted to the two-mode model. However, the mode number $n_{\rm cutoff}=10$ is still far below the one needed in the real-space description \cite{Nagy2008Selforganization}.  

\begin{figure}[ht]
\begin{center}
\includegraphics[width=0.45\textwidth]{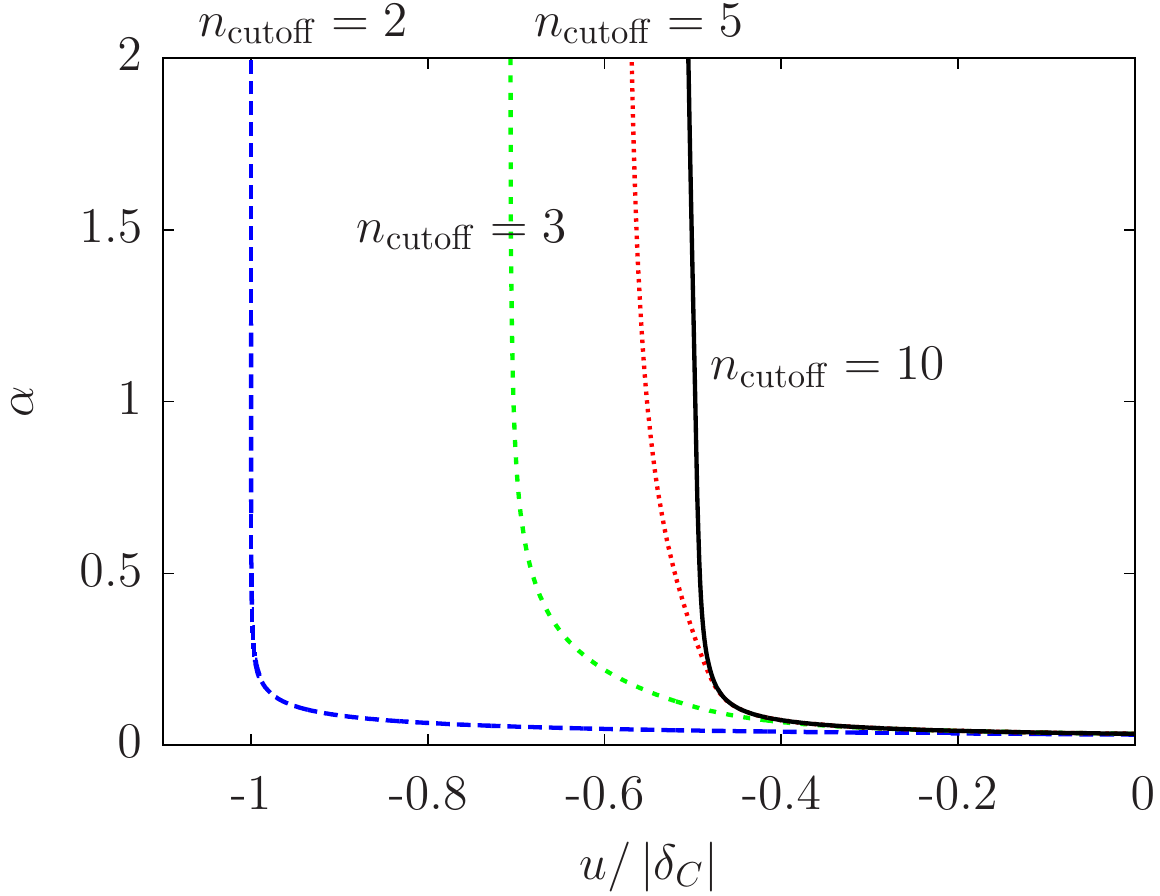}
\end{center}
\caption{The effect of the phase shift term on $\alpha$. The numerical simulation confirms that the effective cavity frequency $\Omega(\gammavec)$ tends to zero at the divergence point. The parameters: $\omega_R=1$, $\delta_C=-100$. The value of $y$ is fixed just above the critical point, $y=11$, $y_{\rm crit}=10$.}
\label{fig:u_alpha}
\end{figure}

The role of higher-order modes is illustrated also in Fig.~\ref{fig:u_alpha}. There is another criticality in the system, of different nature, which occurs when the effective mode frequency $\Omega$, depending on the condensate distribution $\gammavec$ as shown in  Eq.~(\ref{eq:Omega_def}), becomes negative. In this regime there is no stable solution for the coupled atom field and cavity mode system. The effective mode frequency can be tuned by varying $|u|$, which leads to a divergence in $\alpha$. This critical point depends on the cutoff mode index below $n_{\rm cutoff}$. In the two mode case the divergence occurs when $u=\delta_C$, however, the exact result, approached well with the cutoff $n_{\rm cutoff}=10$, is at much smaller $|u|$ because the atoms in higher order modes are allowed to localize much better at the antinodes of the cavity mode function and yield a larger resonance shift.

\subsection{The analysis of the fluctuations}
\label{subsec:fluctuations}

The values of $\alpha$ and $\gammavec$ are known from the numerical solution of \eqref{eq:mean_sol} which also provides for the eigenvectors $\textbf{v}^{(j)}$ associated with the eigenvalues $\lambda_j$  of  $\textbf{M}(\alpha)$. The matrix is symmetric and real, so the eigenvalues and eigenvectors are real. The eigenvectors  form a complete orthonormal basis:
$ \textbf{v}^{{(i)}^T} \cdot \textbf{v}^{(j)} = \delta_{ij}$. We can arrange the eigenvalues in increasing order: the smallest one is $\lambda_0=\mu$, and the corresponding eigenvector is $\textbf{v}^{(0)}=\gammavec$.

Let us first decouple the atomic modes interacting via the second term of \eqref{eq:K2}. The eigenvectors put into the columns of a matrix,
\begin{equation} \label{eq:O_def}
\textbf{O} =
 \left( {\begin{array}{*{20}{c|c|c|c }}
   {} & {} & {} & {}   \\
   \, \textbf{v}^{(0)} \, & \, \textbf{v}^{(1)} \, & \, \textbf{v}^{(2)} \, & \, {\ldots} \, \\
   {} & {} & {} & {}   \\
\end{array}} \right) \;,
\end{equation}
define the orthogonal transformation, $\textbf{O}^T \cdot \textbf{O} = \textbf{O} \cdot \textbf{O}^T =\textbf{I}$, which leads to the bosonic modes
\begin{equation} \label{eq:b_def}
\opvec{b}= \textbf{O}^T \cdot \opvec{c} \; .
\end{equation}
Inversely, 
\begin{equation} \label{eq:b_def_inv}
\opvec{c}= \textbf{O} \cdot \opvec{b} = \textbf{v}^{(0)} \cdot \hat{b}_0 + \textbf{v}^{(1)} \cdot \hat{b}_1 + \textbf{v}^{(2)} \cdot \hat{b}_2 + \ldots \; ,
\end{equation}
Since $\textbf{v}^{(0)}=\gammavec$ the $\hat{b}_0$ mode describes the fluctuations parallel to the condensate. 
Subsequently the $\hat{b}_1$, $\hat{b}_2$, ... modes describe orthogonal excitations.

The grand canonical Hamiltonian simplifies to 
\begin{multline} \label{eq:K_ab}
\hat{K}^{(2)}= \, \Omega(\gammavec) \, \on{\hat{a}} +  \sum_{j=0}^{\infty} \, \left( \lambda_j-\mu \right) \on{\hat{b}_j} \\
+ \frac{1}{2} \sum_{j=0}^{\infty} \, g_j \left( \hat{a}^\dag + \hat{a} \right) \left( \hat{b}_j^\dag + \hat{b}_j \right) \; ,
\end{multline}
where $\textbf{g}= \textbf{O}^T \cdot \textbf{M}'(\alpha) \, \gammavec$. 
The frequency of the $\hat{b}_0$ mode is zero, since $\lambda_0 =\mu$. This zero-mode is the Goldstone mode resulting from the $U(1)$ symmetry breaking imposed by the choice of real mean field $\gammavec$.  In the two-dimensional phase space of the Goldstone mode, the quadrature $\hat r \equiv \tfrac{1}{2}(\hat{b}_0^\dag+\hat{b}_0)$ is parallel with the condensate and is a constant of motion (commutes with the above $\hat{K}^{(2)}$).  The orthogonal quadrature $\sqrt{N_c} \,\hat \phi \equiv i(\hat{b}_0^\dag-\hat{b}_0)/2$ corresponds to phase fluctuations of the condensate \cite{Lewenstein1996Quantum}. It obeys the equation of motion
\begin{equation}
\frac{d}{dt} \hat\phi = - \frac{g_0}{2 \sqrt{N_c}} \left( \hat{a}^\dag + \hat{a} \right)\,,
\end{equation}
\begin{equation}
\frac{d^2}{dt^2} \, \hat{\phi} (t) = - i\, \frac{g_0 \; \Omega(\gammavec)}{2 \, \sqrt{N_c}} \; \left(\hat{a}^\dagger -\hat{a}\right) \,.
\end{equation}
Let $\Delta t$ be a small time interval on the timescale of the variation of the condensate phase. $\hat{\phi} (\Delta t)$ can then be well approximated by the second-order Taylor-expansion which includes the above two time derivatives. 
The growth of the phase fluctuations is characterized by $\ev{\hat{\phi}^{\, 2} (\Delta t)}$. Since $\hat{\phi} (0)$ is not correlated neither with $\hat{a} (0)$ nor with $\hat{a}^\dagger (0)$, the only non-vanishing term up to second order is
\begin{multline}
\ev{\hat{\phi}^{\, 2} (\Delta t)} = \ev{\hat{\phi}^{\, 2} (0)} + \frac{g_0^2}{4 \, N_c} \ev{\left(\hat{a}^\dagger (0) + \hat{a} (0)\right)^{2}} \cdot {\Delta t}^2 \\ +O({\Delta t}^3) \; ,
\end{multline} 
where $\ev{\left(\hat{a}^\dagger (0) + \hat{a} (0)\right)^{2}}$ will be given later by \eqref{eq:xp_def} and \eqref{eq:korr_xp}. However,  this expectation value is close to 1 except for a small vicinity of the critical point where it diverges. In the thermodynamic limit $N_c \rightarrow \infty$ the phase undergoes then a free expansion with characteristic time scale about $\frac{\pi \sqrt{N_c}}{g_0}$, which is the far longest time scale.  Note also that, below threshold $g_0=0$ so the phase fluctuations are not growing at all.

From now on, we neglect these fluctuations and completely eliminate the dynamics of the mode $\hat{b}_0$. This approximation renders the condensate to be a classical background field similar to the external laser pump field which was described by the real parameter $\eta_t$ in the model.
This step is equivalent to projecting the atomic excitation space to the one orthogonal to the condensate, 
$\opvec{c} \rightarrow \opvec{c}_\perp = \opvec{c} - \gammavec (\gammavec^T \opvec{c}) $, as described in \cite{Castin2001BoseEinstein}.

\begin{figure}[ht!]
\begin{center}
\includegraphics[width=0.45\textwidth]{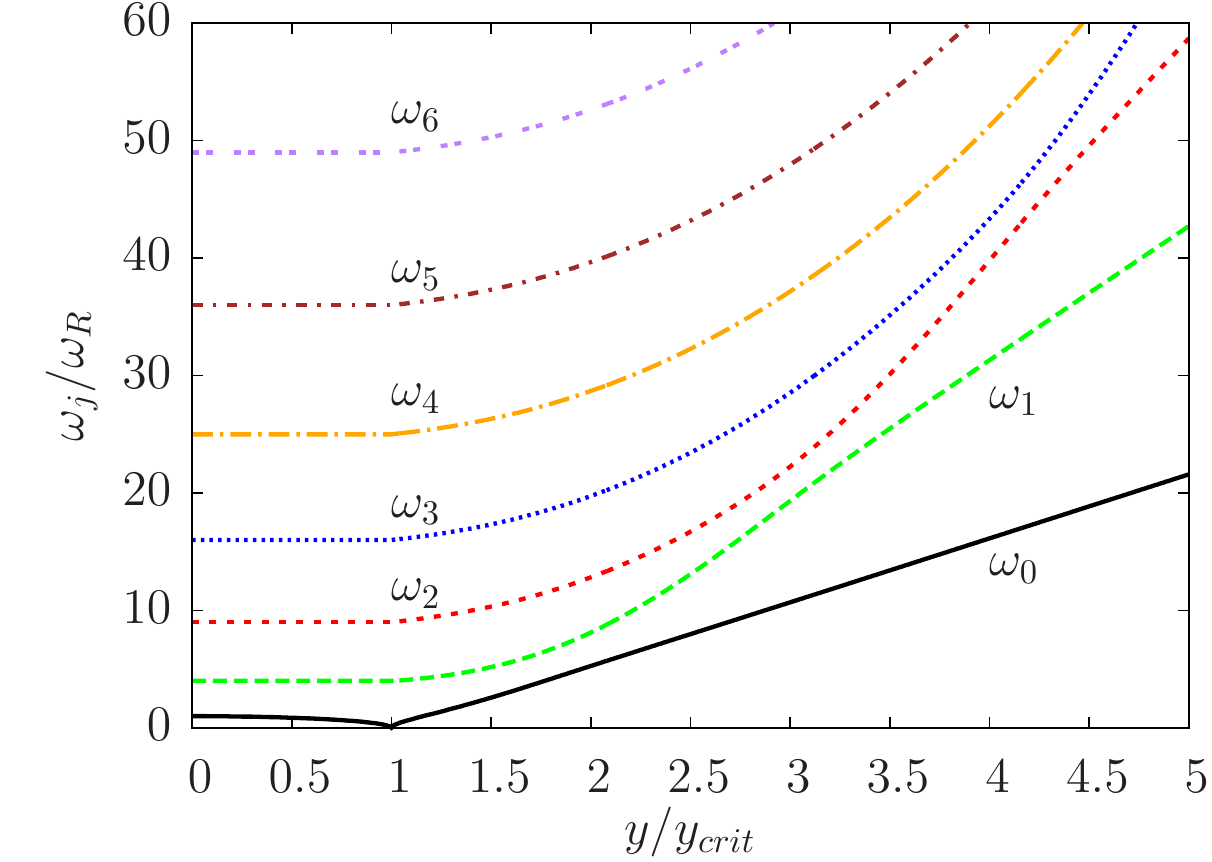}
\end{center}
\caption{The $\omega_j$ eigenfrequencies of the system. For $y=0$ the well-known  $\omega_j= j^2 \cdot \omega_R$ spectrum of an ideal gas in a box is rendered. On increasing $y$, the lowest eigenvalue tends to zero at the phase transition point.  The parameters are the same as in Fig.~\ref{fig:BecDistrib}.}
\label{fig:spectrum}
\end{figure}

In the following, we perform a Bogoliubov-trans\-for\-ma\-tion on $\left( \hat{a} \, , \, \hat{a}^\dag \, , \, \opvec{b} \, , \, \opvec{b}^{\, \dag} \right)$ in order to define the independent quasiparticle modes $\left( \opvec{d} \, , \, \opvec{d}^{\, \dag} \right)$, which combine excitations of the atom field and the electromagnetic field. Let us introduce the quadrature amplitudes
\begin{subequations} \label{eq:xp_def}
\begin{gather}
\hat{x}_0= \frac{1}{\sqrt{2 \, \Omega }} \left( \hat{a}^\dag + \hat{a} \right) \\
\hat{p}_0= i \, \sqrt{\frac{\Omega}{2}} \left( \hat{a}^\dag - \hat{a} \right) \\
\hat{x}_j= \frac{1}{\sqrt{2 ( \lambda_j -\mu) }} \left( \hat{b}_j^\dag + \hat{b}_j \right) \\
\hat{p}_j= i \, \sqrt{\frac{\lambda_j -\mu}{2}} \left( \hat{b}_j^\dag - \hat{b}_j \right) \; ,
\end{gather}
\end{subequations}
where $j\in\{1,2,3,...\}$. Note that the $\left(\hat{x}_0 \, , \, \hat{p}_0 \right)$ quadratures are related to $\hat{a}$, not to $\hat{b}_0$.
The quadratures obey the usual canonical commutation relations:
\begin{equation}  \label{eq:comm_xp}
\comm{\hat{x}_k}{\hat{p}_l}=i \, \delta_{kl} \; ,
\end{equation}
and all other commutators vanish.
Using vector notations,
$\opvec{x}= \left(\hat{x}_0 \, , \, \hat{x}_1 \, , \, \hat{x}_2 \, , \, ... \right)^T \;$ and  
$\opvec{p}= \left(\hat{p}_0 \, , \, \hat{p}_1 \, , \, \hat{p}_2 \, , \, ... \right)^T$, the grand canonical Hamiltonian can be expressed in terms of the quadratures as
\begin{equation}
\hat{K}^{(2)}= \frac{1}{2} \; \opvec{p}^T \, \opvec{p} \; + \; \frac{1}{2} \; \opvec{x}^T \, \textbf{S} \, \opvec{x} \; + \text{const} \; ,
\end{equation}
where c-numbers were omitted and the $\textbf{S}$ kernel matrix is:
\begin{equation}
\textbf{S}= 
\left( {\begin{array}{*{20}{c}}
   {{\Omega ^2}} & {{{\tilde g}_1}} & {{{\tilde g}_2}} & \cdot & \cdot  \\
   {{{\tilde g}_1}} & {{{\left( {{\lambda _1} - \mu } \right)}^2}} & {} & {} & {}  \\
   {{{\tilde g}_2}} & {} & {{{\left( {{\lambda _2} - \mu } \right)}^2}} & {} & {}  \\
   \cdot & {} & {} & \cdot & {}  \\
   \cdot & {} & {} & {} & \cdot  \\
\end{array}} \right)\; ,
\end{equation}
with the off-diagonal elements $\tilde{g}_j = g_j \cdot \sqrt{\Omega (\lambda_j -\mu )}$.
It follows that the Bogoliubov transformation amounts to  the diagonalization of the real symmetric matrix $\textbf{S}$. 
Let $\textbf{U}$ be the orthogonal matrix comprising the eigenvectors of $\textbf{S}$ as its colums ($ \textbf{U}^T \cdot \textbf{U} = \textbf{U} \cdot \textbf{U}^T = \textbf{I}$). The canonical transformation, 
\begin{equation} \label{eq:xp_trafo}
\opvec{x}= \textbf{U} \cdot \opvec{X} \,, \quad \opvec{p} = \textbf{U} \cdot \opvec{P} \,, 
\end{equation}
leads to
\begin{equation}
\hat{K}^{(2)}= \frac{1}{2} \, \sum_{j=0}^{\infty} \, \left( \hat{P}_j^2 \; + \; \omega_j^2 \, \hat{X}_j^2  \right) \; + \text{const} \;,
\end{equation}
where $\omega_j^2$ are the real eigenvalues of the matrix $\textbf{S}$. This Hamiltonian describes independent harmonic oscillators associated with bosonic quasiparticles with $\omega_j$ eigenfrequencies. The annihilation and creation operators of the quasiparticles are
\begin{align} \label{eq:d_def}
\hat{d}_j&= \sqrt{\frac{\omega_j}{2}} \; \hat{X}_j + \frac{i}{\sqrt{2 \, \omega_j}} \; \hat{P}_j \;, \\
\hat{d}_j^{\, \dag}&= \sqrt{\frac{\omega_j}{2}} \; \hat{X}_j - \frac{i}{\sqrt{2 \, \omega_j}} \; \hat{P}_j \; ,
\end{align}
where $j\in\{0,1,2,...\}$, then the
Hamiltonian is
\begin{equation} \label{eq:K_d}
\hat{K}^{(2)}= \sum_{j=0}^{\infty} \, \omega_j \; \hat{d}_j^{\, \dag} \, \hat{d}_j  \; + \text{const} \; .
\end{equation}
The spectrum, the set of the eigenvalues $\omega_j$, is plotted in Fig.~\ref{fig:spectrum} as a function of the pump strength parameter $y$.
The calculation relies on a number of modes $n_{\rm cutoff} = 10$ which reproduces the exact result \cite{Nagy2008Selforganization}.

\subsection{Fluctuations in the normal phase}
\label{subsec:normal_phase_fluctuations}

We apply the general results to the normal phase of the system described by 
$\alpha=0$ and $\gammavec= \left( 1 \, , \, 0 \, , \, 0 \, , \, ... \right)^T$. So, there is no coherent mean optical field in the resonator, and the condensate is homogeneous. 
Then \eqref{eq:Omega_def} and \eqref{eq:M(alpha)_def} gives $\Omega(\gamma)=-\delta_C$ and $\textbf{M}(\alpha=0)=\omega_R \, \textbf{M}^{(0)}$, respectively. This matrix is already diagonal: we can read out that $\mu=0$ and $\lambda_n = n^2 \cdot \omega_R$. The orthogonal transformation in \eqref{eq:b_def} is the trivial one: $\textbf{O}=\textbf{I}$ and $\opvec{b}=\opvec{c}$.
The vector of the coupling constants between $\hat{a}$ and $\opvec{c}$ is: $\textbf{g}= \textbf{O}^T \cdot \textbf{M}'(\alpha) \, \gammavec =y \cdot \left( 0 \, , \, 1 \, , \, 0 \, , \, 0 \, , \, ... \right)^T$. 
This is an important result: in the normal phase $\hat{a}$ is coupled only to $\hat{c}_1$ and the coupling constant is  $g_1=y$. This means that the two-mode model \cite{Nagy2010DickeModel} is exact in the  normal phase, and the Hamiltonian \eqref{eq:K_ab} is simply 
\begin{multline} \label{K_a,c1_csat}
\hat{K}^{(2)}= \, -\delta_C \; \on{\hat{a}} \; + \; \omega_R \,  \sum_{n=1}^{\infty} \, n^2 \, \on{\hat{c}_n}
\\ + \frac{1}{2} \, y \left( \hat{a}^\dag + \hat{a} \right) \left( \hat{c}_1^\dag + \hat{c}_1 \right)\;¬†.
\end{multline}
The $\textbf{S}$ matrix is
\begin{equation}
\textbf{S}= 
\left( {\begin{array}{*{20}{c}}
   {\delta_C^2} & {y \cdot y_{\rm crit}} & {} & {} & {}  \\
   {y \cdot y_{\rm crit}} & {\omega_R^2} & {} & {} & {}  \\
   {} & {} & {4 \, \omega_R^2} & {} & {}  \\
   {} & {} & {} & {9 \, \omega_R^2} & {}  \\
   {} & {} & {} & {} & \cdot  \\
\end{array}} \right) \;,
\end{equation}
with the modified coupling constant $\tilde{g}_1=y \cdot y_{\rm crit}$, where $y_{\rm crit} = \sqrt{-\delta_C \, \omega_R}$ is the critical point of the phase transition in the two mode model as shown in \cite{Nagy2010DickeModel}. Diagonalization of the first block leads to the non-trivial eigenvalues
\begin{equation}
\omega_{\pm}^2= \frac{\delta_C^2+ \omega_R^2}{2} \pm \sqrt{ \left(\frac{\delta_C^2-\omega_R^2}{2}\right)^2 + \delta_C^2 \, \omega_R^2 \, \frac{y^2}{y_{\rm crit}^2} } \; , 
\end{equation}
The frequency $\omega_{-}$ goes to zero at $y=y_{\rm crit}$, which is then the phase transition point. It follows that the phase transition occurs at the point $y=y_{\rm crit}$ even in the multimode model. 

\section{Analysis of the ground state fluctuations in arbitrary phase}
\label{sec:ground_state_analysis}

Our next goal is to express the ground state of the Hamiltonian \eqref{eq:K2} in terms of the Fock space of the operators $\hat{a}$ and $\opvec{c}$ which have clear physical meaning. The ground state of the Hamiltonian \eqref{eq:K_d}  is simply the vacuum state of the $\opvec{d}$ operators. However, since the Bogoliubov transformation in Eq.~(\ref{eq:xp_trafo}) mixes creation and annihilation operators, the ground state contains photonic and motional excitations, moreover, it will be an entangled state. As follows from the Hamiltonian in \eqref{K_a,c1_csat}, below threshold the ground state is the two-mode squeezed vacuum.

We can use  the \emph{Wigner-function} to fully describe the ground state \cite{Hillery1984Distribution}.  We make use of the fact that the Wigner-function associated with the ground state of a bilinear Hamiltonian is always a multivariate Gaussian distribution which, when  centered at the origin, is fully determined by its covariance matrix \cite{Braunstein2005Quantum}. To obtain the Wigner-function, we have to calculate then the symmetrically ordered covariance matrix.

We can start from the correlations between the $(\opvec{X},\opvec{P})$ quadratures pertaining to the independent quasi-particles, which are
\begin{subequations} \label{eq:korr_XP}
\begin{gather}
\ev{ \hat{X}_k \, \hat{X}_l }= \frac{1}{2 \, \omega_k} \, \delta_{kl} \\
\ev{ \hat{P}_k \, \hat{P}_l }= \frac{\omega_k}{2} \, \delta_{kl} \\
\ev{ \szim{ \hat{X}_k \, \hat{P}_l }} = 0 ,
\end{gather}
\end{subequations}
where $\szim{ ...}$ denotes symmetric ordering. 

Now we apply the \eqref{eq:xp_trafo} transformation to determine the covariance matrix of the $(\opvec{x},\opvec{p})$ quadratures associated with the photonic and the atomic motional excitations:
\begin{subequations}  \label{eq:korr_xp}
\begin{gather}
\ev{ \hat{x}_k \, \hat{x}_l } =\frac{1}{2} \sum_{j =0}^{\infty} U_{kj} \, U_{lj} \cdot \frac{1}{\omega_j} \\
\ev{ \hat{p}_k \, \hat{p}_l } =\frac{1}{2} \sum_{j =0}^{\infty} U_{kj} \, U_{lj} \cdot \omega_j \\
\ev{ \szim{ \hat{x}_k \, \hat{p}_l }} = 0
\end{gather}
\end{subequations}
%

\begin{figure}[ht]
\begin{center}
\includegraphics[width=0.45\textwidth]{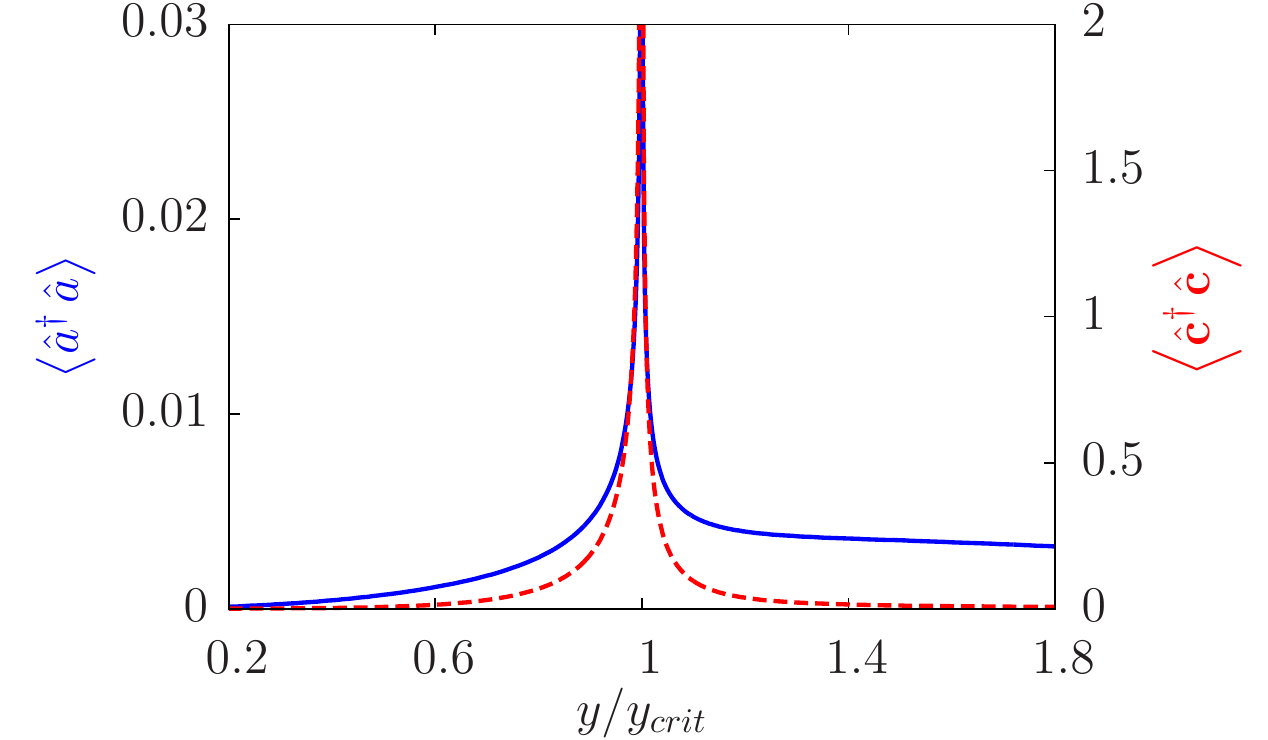}
\end{center}
\caption{The number of the incoherent photons $\ev{\on{\hat{a}}}$ (solid line) and the expectation value of the total number of atoms outside the condensate $\ev{\on{\opvec{c}}}$ (dashed line) near the phase transition point.
The parameters: $\omega_R=1$, $\delta_C=-100$, $u=-20$, $y_{\rm crit}=10$, $n_{\rm cutoff}=3$.}
\label{fig:IncoherentPhotons}
\end{figure}
As the application of the above results, let us calculate the number of incoherent photons in the ground state.
From \eqref{eq:xp_def} and \eqref{eq:korr_xp} it follows that
\begin{equation}
\ev{\on{\hat{a}}} \; = \; \frac{1}{4} \, \sum_{j=0}^{\infty} \, U_{0j}^2 \, \left( \frac{\omega_j}{\Omega} + \frac{\Omega}{\omega_j} -2 \right) \; .
\end{equation}
This expression is numerically evaluated and plotted as a function of the pump strength parameter $y$ in Fig.~\ref{fig:IncoherentPhotons}. With a similar calculation, just performing the replacements $0 \rightarrow k$ and $\Omega \rightarrow \lambda_k- \mu$, we can get the number of atoms in each of the $\hat{b}_k$ modes
\begin{equation}
\ev{\on{\hat{b}_k}} \; = \; \frac{1}{4} \, \sum_{j=0}^{\infty} \, U_{kj}^2 \, \left( \frac{\omega_j}{\lambda_k-\mu} + \frac{\lambda_k-\mu}{\omega_j} -2 \right) \; ,
\end{equation}
where $k \in \{1,2,3,...\}$. By summing these terms, we can get the total number of atoms outside the condensate:
\begin{multline} \label{teljes_atomszam_kondenzatumon_kivul}
\ev{\on{\opvec{c}}}=\ev{\on{\opvec{b}}} 
= \sum_{k=1}^{\infty} \, \ev{\on{\hat{b}_k}} \\ = 
 \frac{1}{4} \, \sum_{k=1}^{\infty} \, \sum_{j=0}^{\infty} \; U_{kj}^2 \; \left( \frac{\omega_j}{\lambda_k-\mu} + \frac{\lambda_k-\mu}{\omega_j} -2 \right) \; ,
\end{multline}
which is plotted in Fig.~\ref{fig:IncoherentPhotons} as a function of the coupling parameter $y$. Note that the index associated with $\hat{c}_k$ runs from 0, while the index associated with $\hat{b}_k$ runs from 1. Finally, the number of atoms in the $\hat{c}_k$ modes, which are associated with the spatial harmonic functions, can be obtained by use of the transformation rule \eqref{eq:b_def_inv}, the inverse of the formulae \eqref{eq:xp_def} and the covariance matrix \eqref{eq:korr_xp}:
\begin{multline}
\ev{\on{\hat{c}_n}} =  \sum_{k,l=1}^{\infty} \, O_{nk} \, O_{nl} \, \ev{\hat{b}_k^\dag \, \hat{b}_l } \\
 =  \, \frac{1}{4} \, \sum_{k,l=1}^{\infty} \; \sum_{j=0}^{\infty} \; O_{nk} \; O_{nl} \; U_{kj} \; U_{lj} \; \cdot \\
\cdot \Biggl( \frac{\sqrt{(\lambda_k-\mu)(\lambda_l-\mu)}}{\omega_j} \; + \; \frac{\omega_j}{\sqrt{(\lambda_k-\mu)(\lambda_l-\mu)}}\\
- \sqrt{\frac{(\lambda_k-\mu)}{(\lambda_l-\mu)}} - \sqrt{\frac{(\lambda_l-\mu)}{(\lambda_k-\mu)}}  \Biggr)
\end{multline}
Starting from this formula, and using the orthogonality of the matrix $\textbf{O}$, the total number of atoms outside the condensate in \eqref{teljes_atomszam_kondenzatumon_kivul} can be verified. 

\section{Entanglement in the ground state}
\label{sec:ground_state_entanglement}

Let us partition the system to the cavity mode and another part including all the motional modes of the atom field. The ground state of the system $\ket{\psi_\text{g}}$, which is the vacuum of the quasiparticles defined by the operators $\hat{d}_k$ ($k=0,1,\ldots$), exhibits bipartite entanglement between the radiation and the matter wave fields. This is similar to the entanglement occurring in the Dicke-model \cite{Emary2003Chaos,Lambert2004Entanglement,Buzek2005Instability}. The entanglement can be simply measured by  the \emph{Neumann-entropy} of the cavity subsystem $\cl{C}$,
\begin{equation}
S_{\cl{C}} = - {Tr}_{\cl{C}} \left( \hat{\rho}_{\cl{C}} \cdot \ln \hat{\rho}_{\cl{C}} \right) \;¬†,
\end{equation}
where  the reduced density matrix of the cavity mode is obtained by tracing the total density matrix $\hat{\rho}=\ket{\psi_\text{g}} \bra{\psi_\text{g}}$ over the atomic subsystem $\cl{A}$, i.e., $\hat{\rho}_{\cl{C}}= Tr_{\cl{A}} \left( \hat{\rho} \right)$.  Another entanglement measure, which can be used, is the \emph{linear entropy},
\begin{equation} \label{lin_ent}
S_{\cl{C}, \, lin} = 1 - {Tr}_{\cl{C}} \left( \hat{\rho}_{\cl{C}}^2 \right)\;.
\end{equation}
Both entropies have zero value precisely if $\hat{\rho}_\cl{C}$ is pure. For mixed states, they are positive, however, $S_{\cl{C}}$ can have an arbitrarily large value, while the linear entropy is bounded, $S_{\cl{C}, \, lin} \leq 1$.

\begin{figure}[t!]
\begin{center}
\includegraphics[width=0.45\textwidth]{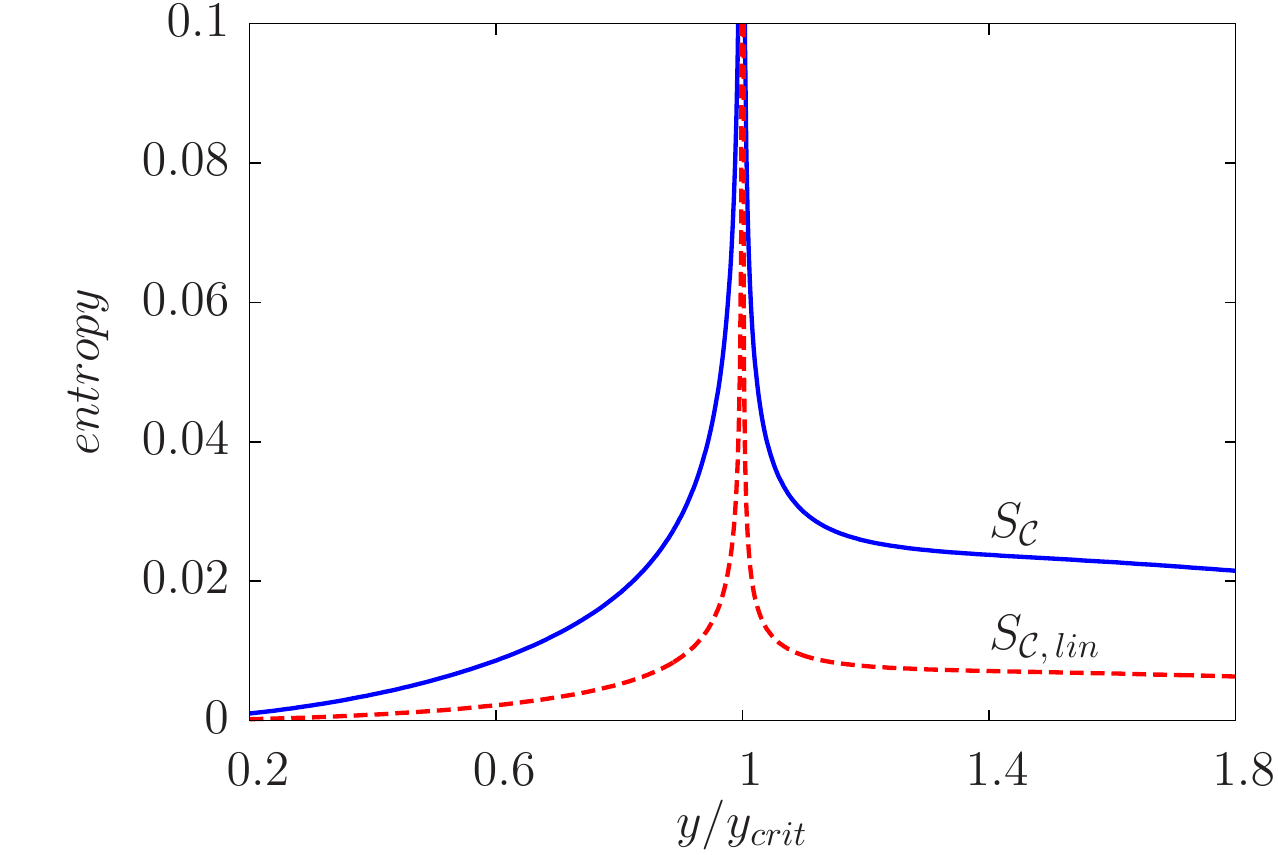}
\end{center}
\caption{\label{fig:entropy} The bipartite entanglement measures, the Neumann-entropy $S_{\cl{C}}$ and the linear entropy $S_{\cl{C}, \, lin}$ as a function of the coupling parameter $y$. 
Parameters: $\omega_R=1$, $\delta_C=-100$, $u=-20$, $y_{crit}=10$, $n_\text{max}=3$.}
\end{figure}
Let us introduce the parameter $\chi$,
\begin{equation}
\label{eq:chi_def}
\chi = \sqrt{- \evN{ \left( \hat{a}^\dag + \hat{a} \right)^2} \evN{ \left( \hat{a}^\dag - \hat{a} \right)^2} } = 2 \cdot \sqrt{ \ev{\hat{x}_0^2} \ev{ \hat{p}_0^2} } \,,
\end{equation}
which leads to a simple form of the entanglement measures in the ground state of the total system \cite{Barthel2006Entanglement,Vidal2007Entanglement},
\begin{subequations} \label{S_chi}
\begin{gather}
S_{\cl{C}} = \frac{\chi+1}{2} \cdot \ln \left( \frac{\chi+1}{2} \right) - \frac{\chi-1}{2} \cdot \ln \left( \frac{\chi-1}{2} \right) \\
S_{\cl{C}, \, lin} = 1 - \frac{1}{\chi}
\end{gather}
\end{subequations}
It can be seen that both functions are strictly increasing with increasing $\chi$. It follows from the definition Eq.~(\ref{eq:chi_def}) and from the Heisenberg uncertainty relation that $\chi \in [+1,+\infty)$, and the minimum $\chi=1$, that is, a pure ground state corresponds to a minimal uncertainty state of the field mode. Here we exploited the relation $\ev{\hat{x}_0}=\ev{\hat{p}_0}=0$, so that the quadrature variances are $\Delta x_0 = \sqrt{ \ev{\hat{x}_0^2}}$ and $\Delta p_0 = \sqrt{ \ev{\hat{p}_0^2}}$. The entropies are plotted in Fig.~(\ref{fig:entropy}), and they exhibit singularity at the critical point, similarly to the incoherent photon and atom numbers in Fig.~\ref{fig:IncoherentPhotons}.

\section{Conclusion}

In this paper we presented a ground-canonical mean field theory for the coupled system of a single cavity mode and many motional modes of an ultracold atom field. We identified all the necessary approximations to separate the mean field from the fluctuations. The theory has been applied to describe the self-organization phase transition \cite{Nagy2008Selforganization}.  We showed that below, in the normal phase, and in the vicinity of the critical point the two-mode approximation for the atomic motion holds, therefore the system shows indeed an analogy with the Dicke model of superradiant quantum phase transition \cite{Nagy2010DickeModel}. Well above threshold the multi-mode expansion is needed, however, our approach leads to mean-field equations which have to be solved numerically.  We calculated the number of incoherent photons and populations in the higher excited motional modes, as well as the amount of entanglement between the matter and radiation fields. This approach is suitable and will be used in the future to deal with other type of multimode systems, for example, the matter wave field coupled to the radiation field in a degenerate confocal resonator which shows a rich phase diagram \cite{Gopalakrishnan2009Emergent,Gopalakrishnan2010Atomlight}.

\begin{acknowledgement}
This work was supported by the National Office for Research and Technology under the contract ERC\_HU\_09 OPTOMECH, and the European Science Foundation's EuroQUAM project {\em Cavity-Mediated Molecular Cooling}. G.Sz. also acknowledges funding from the Spanish MEC projects TOQATA (FIS2008-00784), QOIT (Consolider Ingenio 2010), ERC Advanced Grant QUAGATUA and EU STREP NAMEQUAM.
\end{acknowledgement}


\end{document}